\documentclass{aa} 
\usepackage{natbib}
\usepackage{graphicx}
\usepackage{txfonts}
\begin{document}
   \title{Star formation history of galaxies from z=0 to z=0.7}

   \subtitle{A backward approach to the evolution of  star-forming galaxies.}
%%%%A secular  evolution for  star-forming galaxies more massive than $\rm 10^{10} M_{\odot}$}

   \author{V. Buat
          \inst{1}
          \and
           S. Boissier\inst{1}
\and
D. Burgarella\inst{1} \and   T. T. Takeuchi \inst{2}  \and E. Le Floc'h \inst{3} \and D. Marcillac \inst{4} \and J. Huang \inst{5} \and M. Nagashima \inst{6} \and M. Enoki \inst{7}}

   \offprints{V. Buat}

   \institute{Observatoire Astronomique Marseille Provence, Laboratoire d'Astrophysique de Marseille,
BP8, 133761 Marseille cedex 12, France\\
              \email{veronique.buat@oamp.fr}\\
          \and
Institute for Advanced Research, Nagoya University,
Furo-cho, Chikusa-ku, Nagoya 464-8601, Japan\\
 \and
Spitzer fellow, Institute for Astronomy, University of Hawaii, 2680 Woodlawn Drive, Honolulu, HI 96822\\
\and
Steward Observatory, University of Arizona, 933 North Cherry Avenue, Tucson, AZ 85721, USA\\
\and
Harvard-Smithsonian Center for Astrophysics, 60 Garden Street, Cambridge, MA 02138\\
\and
Faculty of Education, Nagasaki University, Nagasaki 852-8521, Japan\\
\and
Faculty of Bussiness Administration, Tokyo Keizai University,
1-7-34, Minami-cho, Kokubunji, Tokyo, 185-8502, Japan
}

   \date{}

% \abstract{}{}{}{}{} 
% 5 {} token are mandatory
 
  \abstract
  % context heading (optional)
  % {} leave it empty if necessary  
   {}
  % aims heading (mandatory)
   { We investigate whether the mean star formation activity of  star
     forming galaxies from z=0 to z=0.7 in the GOODS-S field can be reproduced by simple
     evolution models of these systems. In this case, such models
     might be used as first order references for
%a zero point (reference) 
     studies at higher z
      to decipher when and to what extent a secular evolution
     is sufficient  to explain the star formation history in galaxies.}
  % methods heading (mandatory)
   {We selected star-forming galaxies at z=0 and at z=0.7 in IR and in
     UV  to have access to all the  recent star formation.
     We focused on galaxies with a stellar mass ranging between
     $10^{10}$ and $10^{11}$ M$_{\odot}$ for which the results are not
     biased by the selections. We compared the data to 
     chemical evolution models developed for spiral galaxies and
     originally built to reproduce the main characteristics of the
     Milky Way and  nearby spirals without fine-tuning them for the present analysis.}
  % results heading (mandatory)
   {We find a shallow decrease in the specific star formation rate
     (SSFR)  when the stellar mass increases. The evolution of the SSFR  characterizing both UV and IR
      selected galaxies from z=0 to z=0.7 is consistent with the
     models built to reproduce the present spiral galaxies. There is no need
      to strongly modify of the physical
     conditions in  galaxies to explain the average evolution of their star formation
     from z=0 to z=0.7. We use the models to 
     predict the evolution of the star formation rate and the metallicity
     on a wider range of redshift and we compare these predictions with the results of semi-analytical models}. 
     
   {}
\keywords{galaxies: evolution-galaxies: stellar content-infrared: galaxies-ultraviolet: galaxies
               }

   \maketitle
%
%________________________________________________________________

\section{Introduction}

A lot of recent studies have explored the relation between the stellar
mass and the star formation rate (SFR) in galaxies at different
redshifts. It has been shown that star formation  critically depends on
galaxy mass both at low and high redshift \citep[e.g.][and reference
therein]{gavazzi02,brinchmann04,feulner}. The general process called
``downsizing'' \citep{cowie} is now commonly accepted: it can be
summarized by an early and rapid formation of massive galaxies, whereas
low-mass systems evolve more smoothly, being still actively forming
stars at z=0. The galaxies are usually subdivided into active star-forming and quiescent systems within which star formation has been
quenched.   When only star-forming galaxies are concerned, the
  specific star formation rate (star formation rate divided by stellar
  mass) seems to exhibit a flatter distribution as a function of the
  stellar mass than when the whole population of galaxies is accounted
  for \citep[e.g.][]{daddi07,elbaz07,iglesias07}.

 Physical processes that might be at the origin
of or might influence the quenching of the star formation in massive
galaxies have been extensively discussed \citep[e.g.][and references
therein]{dekel06, bundy05, bundy06}. Conversely there are only very few
attempts to propose realistic star formation histories to explain the
trends  that are observed: most of the studies only explore  very crude scenarios
like a constant star formation rate or an instantaneous burst
\citep{feulner}. Recently, \citet{noeske07} have proposed a 
more sophisticated model with an exponential star formation history
whose parameters are mass-dependent: less massive galaxies have a
longer e-folding time and begin their formation more recently.
Their model is based on two parameters: the e-folding time $\tau$ and
redshift of galaxy formation $z_f$,  which are fitted to their data. The
redshift formation $z_f$ is likely to be representative of the bulk of
the star formation: a galaxy with a stellar mass of $\rm \sim 10^{10}
M\odot$ would form at $z_f=1$ against $z_f=3$ for more massive objects
$\rm \sim 10^{11} M\odot$.

In this paper, we follow a different approach. We  start from
a physical model built to reproduce the mean properties of local star-forming disk galaxies. We  analyze how this model, without any
modification or adjustment, can reproduce the evolution of the star
formation observed with z. Practically speaking, we  use a grid of models
predicting the chemical and spectrophotometric evolution of spiral
galaxies. These models were calibrated in the Milky Way
\citep{boissier99} and successfully reproduce the properties of nearby
spiral galaxies \citep{boissier00}. The redshift formation $z_f$ is
taken to be equal to 6 for all the galaxies and represents the time when the
first stars begin to form.

Our work is motivated by the recent
findings that the UV and IR luminosity functions at intermediate
redshifts (up to z $\sim 0.7-1$) are almost dominated by normal spiral
or irregular galaxies
\citep{bell05,wolf05,melbourne,zheng07b, zamojski}. The role of major mergers
does not seem to govern the star formation at these redshifts, one
can  expect those galaxy evolution models that assume  a smooth
evolution to be appropriate for the interpretation of the observations.

In section 2 we  describe the samples of sta-forming galaxies at
z=0 and z=0.7 and the derivation of the main parameters useful to this
study: the star formation rates (SFR) and stellar masses (M$_{\rm star}$).
The selections are performed in rest-frame IR (60 or 15$\mu$m) and UV
(1530$\rm \AA$) in order to be sure to pick up all  star-forming
galaxies at these redshifts. The characteristics of these selections
 are  outlined in section 3. Specific star formation rates (SFRs
divided by M$_{\rm star}$) are analyzed in section 4.  The evolutionary
models are presented in section 5  where we propose a simple analytical
formulation for the SFR history  of each galaxy. In section 6 the models
are compared to the data. We also compare our results to more 
sophisticated (semi-analytical) models and give predictions for higher z.
Section 7 is devoted to the conclusions. \\

Throughout this article, we use the cosmological parameters $H_0 = 70$
km s$^{-1}$ Mpc$^{-1}$, $\Omega_M = 0.3$ and $\Omega_{\lambda} = 0.7$.
All magnitudes are quoted in the AB system.  The IR luminosity
$L_{\rm IR}$ is defined over the wavelength range 8-1000 $\mu$m. The
UV luminosity $L_{\rm UV}$ is defined as $\nu L_{\nu}$. The  luminosities are expressed in solar units with 
$\rm L_\odot = 3.83~10^{33} \rm {erg s^{-1}}$

%__________________________________________________________________

\section{Data sets, SFRs and stellar mass estimates}

Our goal is to study the evolution of star-forming galaxies from the local
universe to intermediate redshift ($z \leq \sim 1$) by comparing observations to the
predictions of models of their evolution.
Therefore we need to build
galaxy samples that are representative of the overall star formation
and we must avoid quiescent galaxies.  The best way to focus on star-forming galaxies is to select them according to their SFR. Newly
formed stars emit most of their light in the UV and a large fraction
of their emission is re-processed in the IR via dust heating. Therefore we
  select the galaxy samples by
using these two wavelength ranges: UV and IR. 

We  use local samples already built from GALEX and IRAS data at
1530 $\rm \AA$ and 60 $\mu$m \citep{buat06}. At intermediate z the deepest
GALEX and SPITZER surveys can be used to build similar samples to 
z=0. Practically speaking,  we  work at z=0.7 to avoid K-corrections. At this
redshift the GALEX near-ultraviolet band at 2310 $\rm \AA$ corresponds to
a 1530 $\rm \AA$ rest frame, the UV wavelength at which galaxies have been selected at z=0 (far-ultraviolet band of
GALEX). The SPITZER/MIPS
observations at 24 $\mu$m correspond to $\sim 15~\mu$m in the
rest frame of galaxies at z=0.7. Although 15 $\mu$m does not directly
correspond to the IRAS 60$\mu$m band, this mid-infrared wavelength
range has been intensively studied, so  we will be able to derive total
infrared luminosities $L_{\rm IR}$ from these mono-wavelength data. 

 Rest-frame, near-infrared (NIR) data will be needed to
estimate stellar masses, whereas the UV/IR selections ensure  we can 
measure the current star formation: at z=0 they were  estimated by \citet{buat06} from 2MASS data. At z=0.7 the IRAC observations at 3.6 $\mu$m correspond  to the K-band rest frame.\\
All the UV data are
corrected for Galactic extinction using the \citet{schlegel} maps and
the Galactic extinction law of \citet{cardelli}. We have gathered the main characteristics of the samples described  below  in Table 1.

\subsection{z=0 samples}

\citet{buat06}  built two samples of galaxies selected at 60
$\mu$m (IR selected sample) and at $1530\rm \AA$ (UV selected sample) with
a very high detection rate at $1530\rm \AA$ and 60 $\mu$m, respectively. A short description of
the samples is made in Table 1. The original samples were flux-limited: the limiting luminosities reported in the table correspond to the faintest bins of the luminosity functions built with these samples \citep[cf. Fig.3 of][]{buat06}.
Here we  use the mean (volume-averaged) trends found from these samples. The SFRs
are estimated by adding the star formation rate measured from the IR
and the observed UV emissions as preconised by \citet{iglesias}
assuming a constant SFR over $10^8$ years. The only modification that we
perform here is to use the initial mass function (IMF) of \citet{kroupa01} since it is
commonly used in recent studies and it is more consistent with the
models used in this paper.  From Starburst99 \citep{leitherer} we obtain:

\begin{eqnarray}
 \log(SFR_{\rm IR})_{\rm M_{\odot}~yr^{-1}} = \log(L_{\rm IR})_{\rm L_{\odot}}-9.97 \\
\log(SFR_{\rm UV})_{\rm M_{\odot}~yr^{-1}} = \log(L_{\rm UV})_{L_{\odot}}-9.69
\end{eqnarray}
We follow \citet{iglesias} to estimate the total SFR:
\begin{eqnarray}
SFR = (1-\eta)\cdot SFR_{\rm IR}+SFR_{\rm UV} 
\end{eqnarray}
where $\eta$ is the fraction of dust emission due to the heating by old
stars and not related to the recent star formation. The introduction
of this factor was found to be mandatory for making consistent star formation
indicators \citep{bell03,hirashita03,iglesias04,iglesias}. Following
\citet{iglesias} we take $\eta=0.3$.

The stellar mass is calculated from H magnitudes as in \citet{buat06}.
We use the \citet{belletal03} M/L$\rm _H$ calibrations adopting a mean B-V
color of 0.6 mag \citep{buat06} and a \citet{kroupa93} IMF. It
corresponds to M/L$\rm _H$=0.58 in solar units. According to
\citet{bell07}, using  a \citet{kroupa01} IMF instead of a
\citet{kroupa93} one would lead to similar stellar masses (within
10$\%$). We have also checked \citep{buat06} that very similar masses
would be obtained using K magnitudes instead of the H ones.\
\begin{table*}%t1
\caption{Description of the samples. At z=0 all the characteristics come from \citet{buat06}, 
the limiting $L_{\rm IR}$ is  estimated from the luminosity at 60 $\mu$m by applying a bolometric correction of 2.5 \citep{buat06}. 
The samples at z=0.7 are fully described in section 2.2 }\label{table1}
\centering
\begin{tabular}{ccc}

\hline 
 &z=0 & z=0.7  \\
\hline
&IRAS $\rm f_{60}>0.6 Jy$&SPITZER/MIPS $\rm f_{24}>0.024 mJy$\\
IR selection&665 galaxies&280 galaxies\\
&$L_{\rm IR}\ge 2.5 \, 10^9 \rm L_{\odot}$&$ L_{\rm IR}\ge 2.2 \, 10^{10} \rm L_{\odot}$\\
\hline
&GALEX $\rm FUV (1530\AA)  \le 17 mag$&GALEX $\rm NUV (2310\AA) \le 25.5 mag$\\
UV selection&656 galaxies&247 galaxies\\
&$L_{\rm UV}\ge 3 \, 10^8 \rm L_{\odot}$&$ L_{\rm UV}\ge 2.5 \, 10^{9} \rm L_{\odot}$\\
\hline
\end{tabular}
\end{table*}
\subsection{z=0.7 samples}

\subsubsection{IR-selected sample}
 We start with the GOODS observations of the   Chandra deep field south (CDFS) \citep{goods}. The source extraction and flux measurements  were performed at 24 $\mu$m  using DAOPHOT \citep{stetson} in a same way as in \citet{lefloch}. The detection limit at 3 $\sigma$ is found to be 0.016 mJy. We have checked the completeness of our data by comparing the slope of the differential counts with models \citep{lagache03, takeuchi01} and data obtained in the HDF-N \citep{chary04}. The data appear to be complete down to $\sim$0.024 mJy. These values  fully agree with those found by  \citet{elbaz07} for the same set of data. We adopt a secure limit of 0.024 mJy to build our sample  \citep{elbaz07}. Cross correlations with COMBO-17 \citep{wolf} were performed to get the redshift of the sources within a tolerance radius of 2" \citep{lefloch}. 88$\%$ of the sources  have a single counterpart in 2". We restrict the final sample to these objects with a single counterpart. The  optical photometry is retrieved from the EIS data \citep{arn}

To build the sample at z=0.7 we selected galaxies in the redshift bin
0.6-0.8. We put a limit in luminosity to be complete in the redshift
range and to avoid volume corrections. This limit is calculated for a
limiting flux of 0.024 mJy at z=0.8, and translates to a total infrared
luminosity $\log(L_{\rm IR})= 10.34 ~(\rm L_{\odot})$ (see below for
the calibration of the 24 $\mu$m flux in $L_{\rm IR}$). 280 sources
are obtained that way (the faintest object has a 24 $\mu$m flux equal
to 0.031 mJy). All these galaxies are detected with IRAC at 3.6
$\mu$m.\\ We must estimate the total infrared luminosity $L_{\rm IR}$
from 8 to 1000 $\mu$m to measure the ``obscured'' SFR. The
extrapolation from the 24 $\mu$m emission alone to the total IR
emission relies on local templates
\citep[e.g.][]{lefloch,bell05,bell07}: tight relations have been found
between the rest frame 12-15 $\mu$m and the total infrared emission of
local galaxies \citep{chary01,ttt1}.   Recent studies based on
SPITZER data suggest that there is not a strong evolution in the IR
 spectral energy distributions for intermediate redshift galaxies \citep{zheng07b}. At z=0.7,
24 $\mu$m corresponds to 14 $\mu$m and a calibration at 15 $\mu$m can
be used. \citet{ttt1} and \citet{chary01} propose such
calibrations. Here we  use the relation of \citet{ttt1} based on
IRAS and ISO data and local templates from \citet{dale01}  to
be consistent with z=0 calculations. Using the \citet{chary01}
relation would lead to a slightly  higher value of $L_{\rm IR}$ by 0.08
$\pm$ 0.01 in log units. \\ 115 galaxies are detected at 2310 $\rm
\AA$ (NUV band from GALEX). At z=0.7 it corresponds approximately to
the rest-frame FUV band of GALEX centered on 1530 $\rm \AA$
\citep[cf][]{buat07}. In the same way, the 3.6 $\mu$m corresponds to a
rest-frame K band. Therefore we can avoid K-corrections.  For the
galaxies detected at 24 $\mu$m, and in UV the SFR is estimated by
combining the UV and IR emissions in a similar way to z=0 (Eq. 3).
 Since galaxies at higher z are more active in star formation than at z=0,
we can expect a lower contribution of old stars to dust heating and
hence a value of $\eta$ lower than 0.3. To check this issue, we  
followed the method of \citet{iglesias04} by comparing SFRs
calculated with Eq. 3 to  those deduced from the UV luminosity alone, 
corrected for dust attenuation using the recipe of \citet{buat05}. The
SFRs were found to be consistent if $\eta$=0; i.e., all the dust heating is
attributed to young stars in these objects. Therefore we adopted 
$\eta$=0 to calculate the SFRs at z=0.7.  For the galaxies not
detected in the UV, the SFR was calculated with the IR luminosity only
(again with $\eta$=0). The contribution of the UV emission at the
detection limit of UV=26.2 mag is negligible in the estimation of the
SFR (see section 4).\\ As for the z=0 samples, the stellar masses were estimated
following \citet{belletal03} who calibrate the M/L ratios as a
function of several colors including those of the SDSS. The observed
3.6 $\mu$m band corresponds to the K band at z=0.7 and  the observed R-I
color from the EIS survey (Johnson-Cousin system) is similar to the u-g
rest-frame color (from the SDSS). Therefore we used the u-g
color-M/L$\rm _K$ relation of \citet{belletal03} for the 305 galaxies
with a R-I color. 18 galaxies have no R-I color so  for them we used the
mean M/L$\rm _K$ obtained for the sample i.e. M/L$\rm _K$=0.46 (solar
units). The R-I color distribution is  discussed in the next
section.

\subsubsection{UV-selected sample}

GALEX \citep{morissey} observed the CDFS for 76 ks in both the FUV
(1530 $\rm \AA$) and the NUV (2310 $\rm \AA$) as part of its deep imaging
survey. The reduction of the data is extensively described in 
\citet{burgarella2}. Very briefly speaking, we used DAOPHOT to perform
PSF-fitting and disentangle close neighbors. The completeness at a
level of 80$\%$ was obtained at NUV=26.2 mag. The data are available
from \citet{burgarella2}. The cross-identification with the COMBO-17
sources is also described in \citet{burgarella2}. About 70\% of the
GALEX sources are identified in COMBO-17 with a strong dependence on
the NUV magnitude. Truncating at NUV=25.5 mag ensures us that more
than 80$\%$ of the GALEX sources are identified. We adopt this
limiting magnitude in the following and we restrict the final sample
to objects with a single counterpart in COMBO-17 ( 90 $\%$ of the UV
sources have more than one counterpart in COMBO-17). As for the IR
selection to avoid volume corrections, we put a limit in
luminosity, which ensures that we  detect all the galaxies brighter than
this limit in the redshift range 0.6-0.8. This limit is calculated for
a magnitude NUV=25.5 at z=0.8.  It corresponds to $\log(L_{\rm UV})= 9.39
~(\rm L_{\odot})$ and  247 galaxies are selected. 48$\%$ (119/247) of these
sources are detected at 24 $\mu$m.  For the undetected ones, we 
adopted an upper limit at 0.016 mJy, which corresponds to the detection
limit at 3 $\sigma$ (section 2.2.1). The IR luminosities $L_{\rm IR}$ were
estimated from the 24 $\mu$m flux in the same way as for the IR-selected sample (section 2.2.1).  The stellar masses were estimated in
the same way as for the IR sample at z=0.7 and 230 galaxies have a R-I
color. For the 17 remaining objects without a R-I color, we used the
mean M/L$\rm _K$ obtained for the sample i.e.  M/L$\rm _K$=0.42 (solar units).
The R-I color distribution will be discussed in the next section.

The star formation rate was estimated as for the IR selection when UV
and 24 $\mu$m fluxes are available. We  again took $\eta=0$; this
choice is also validated by the comparison between the SFRs calculated
with Eq. 3 and $\eta=0$ and those estimated from the UV luminosity
corrected for dust attenuation (as for the IR selection, section
2.2.1). An upper
limit on the SFR was calculated for the galaxies with only an upper limit at 24$\mu$m . This time we could not neglect the
contribution of the IR emission which is dominant, even at the
detection limit level (see next section).

\section{Comparison of the UV and IR selections}
Before  discussing the star formation activity in  both  samples we  analyze which selection (IR or UV) is best suited to our analysis.
 At z=0 \citet{buat06}  showed that the intrinsically brightest galaxies are lost in a UV selection.  Conversely, intrinsically faint galaxies are  hardly detected in IR. Because our present study is  devoted to a comparison between low and higher z samples we do  not discuss the intrinsically faint objects against which we are strongly biased (cf next section). At z=0, intrinsically bright galaxies are rare and the differences between both selections are small \citep{buat06}; 
however, the galaxy population seen either in UV or in IR is known to brighten as z increases,  we must also check the    differences in  the selections at z=0.7.

\subsection{The relative contribution of the IR and UV emissions to the bolometric luminosity of young stars}

At z=0 \citet{buat06}  compared the bolometric luminosities of the galaxies
defined as $L_{\rm bol} = (1-\eta) L_{\rm IR}+L_{\rm UV}$. Briefly summarized,
we showed that the UV luminosity alone is unable to reproduce the
bolometric luminosity even for UV-selected galaxies and that the
combination with the IR is mandatory. Once both luminosities are added
to calculate $ L_{\rm bol}$, a deficiency of very bright objects (in
terms of $L_{\rm bol}$) is observed in the UV selection as compared to
the IR one. Given the low number of such galaxies, these differences
are slight as is shown in the next section, which is  devoted to specific
star formation rates.

\begin{figure}
   \centering
   \includegraphics[angle=-90,width=13cm]{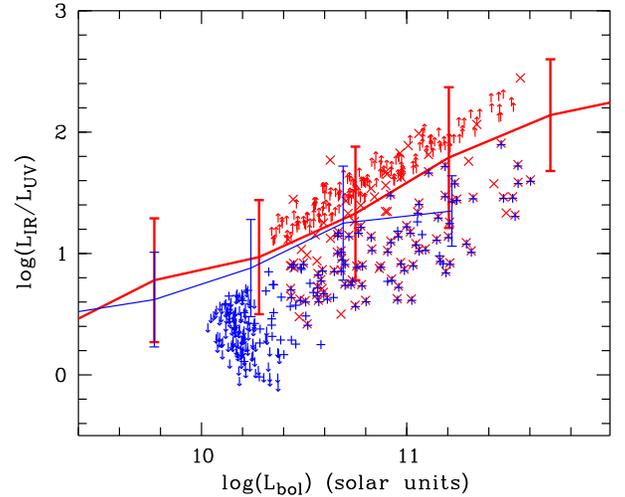}
      \caption{IR to UV luminosity ratio as a function of the bolometric luminosity for both samples at z=0 and z=0.7. At z=0.7 individual data are plotted: UV selection with blue "plus"  for galaxies also detected at 24$\mu$m and  arrows down for galaxies not detected at 24$\mu$m; IR selection with red crosses  for galaxies also detected at UV and arrows up for galaxies not detected at UV. At z=0 volume average trends \citep{buat06} are plotted with solid lines and error bars: the UV selection with a light solid line and the IR selection with a heavy solid line}
              \label{lIR_UV-lbol}
   \end{figure}

In Fig.~\ref{lIR_UV-lbol} we have plotted the IR to UV luminosity
ratio versus $ L_{\rm bol}$ for both samples at z=0.7. It can be seen that the IR luminosity is higher than the UV one for all the galaxies detected at both wavelengths since $L_{\rm IR}/L_{\rm UV}>1$. The galaxies selected in UV and not
detected at 24 $\mu$m exhibit the lowest bolometric luminosities with
$L_{\rm bol} \le 2.5 ~10^{10} \rm L_{\odot}$ ($\log(L_{\rm bol})\le 10.4$). The
galaxies selected in IR and not detected in UV span a very wide range
of bolometric luminosity and hence of SFRs. The average trends found at z=0 are also reported in Fig.~\ref{lIR_UV-lbol}. The samples at z=0 and z=0.7 cover the same range of luminosity. The slight difference found in the distributions of the IR to UV luminosity ratio has been discussed in terms of dust attenuation by \citet{buat07} (see also \citet{burgarella2}).

Some galaxies of the IR selection, detected in UV, do not appear in
the UV-selected sample. The same is true for some galaxies of the UV
selection that have an IR detection but are not included in the IR-selected sample: these ``single'' objects appear in
Fig.~\ref{lIR_UV-lbol} as simple crosses (IR selection) or ``plus''
symbols (UV selection) and represent $30\%$ of both samples. Their
presence is due to the method used to cross-correlate the data: for the IR (resp. UV) selection, the cross-correlation with the UV
(resp. IR) sources is made using the complete list of the UV (resp.
IR) detections before their identification with COMBO 17 objects and any other selection. The truncations
to have complete volume-limited samples account for a half of
the ``single'' objects. The other half
of the IR (resp UV) ``single'' objects have a UV (resp IR) counterpart
 that  has not been identified with a COMBO17 source. Such a rate of non-identification (15 $\%$) is consistent with those quoted by \citet{burgarella2}.

\subsection{Color distributions}

It is well established that a population selected in optical or in NIR
exhibits a bimodal distribution of colors. This result, first obtained
at low z \citep[e.g.][]{baldry}, is also observed at higher z
\citep[e.g.][]{bell04,cooper,bell07,elbaz07}. Blue galaxies are star-forming
objects, whereas the redder ones are quiescent systems. Our galaxies
were selected either in UV or in IR and so are likely to  show 
  active  star formation. For the sake of comparison with
other studies, we can check what sort of galaxies we are dealing with
in terms of optical colors. Unfortunately for us, this check is very
difficult to perform at z=0 since our sample was built from 2000
deg$^2$ not covered by homogeneous optical surveys like the SDSS.
Nevertheless, we can rely on other studies of UV or IR selected
galaxies.  \citet{iglesias07} studied UV-selected galaxies from
z=0.2 to z=0.7. The U-V distribution of their sample remains unimodal
over  the full  redshift range they  analyzed: according to the
classical subdivision into blue and red galaxies, only blue star-forming
galaxies are selected in UV. \citet{goto}  studied the optical
properties of a sample of galaxies selected from the IRAS catalog. The
color distribution that he found is broad and extends from the blue peak to
the red one defined from the SDSS studies.

\begin{figure}
   \centering
   \includegraphics[angle=-90,width=13cm]{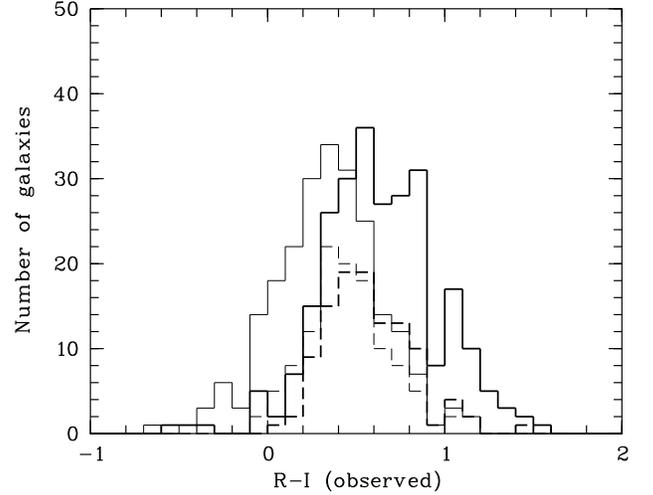}
      \caption{Observed R-I color (AB scale). Heavy  solid line: whole IR selection, heavy  dashed line: IR selected galaxies also detected in UV. Light solid line; whole UV selection, light  dashed line: UV selected galaxies also detected at 24 $\mu$m}
              \label{R_Idist}
   \end{figure}

At z=0.7 the situation is more favorable to our galaxy samples. We
can work with the R-I color from the EIS catalog, which is similar to
the u-g color in the rest-frame of the galaxies. The R-I color is
converted in AB magnitudes according to the conversion formulae of
\citet{arn}. The color distributions are displayed in
Fig.~\ref{R_Idist}. The distinction between red and blue populations
is below and above $R-I\simeq 1$ \citep{elbaz07}.

The red population is under-represented for both selections but the
two distributions clearly differ. As expected in UV we
select preferentially blue galaxies that are  active in star-formation. The
selection looks similar to what is  found by \citet{bundy06} for the DEEP2
survey restricted to galaxies with a SFR larger than 0.2 $\rm M_{\odot}
yr^{-1}$ measured with the [OII] equivalent width. The distribution of
the IR-selected galaxies is broader and shows a tail towards redder
objects, but we do not  see any clear bimodality as reported for example 
for optically selected galaxies (see references above). This agrees with the findings of \citet{bell05} at a similar redshift that the galaxies detected  at 24 $\mu$m are spread over a wide range of optical colors. If we compare, for example, to the distribution of
\citet{elbaz07} it seems that our distribution is broader with a
substantial fraction of our sample between the two peaks defined in
U-B rest-frame (similar to observed R-I at z=0.7). Therefore a
selection at 24 $\mu$m does not exhibit  bimodality, and it also seems 
to be the case at z=0 for galaxies selected from IRAS: IR-selected
galaxies seem to populate the ``green valley'' located between the blue
and red peaks. A  substantial fraction of these galaxies with
intermediate colors exhibit a strong dust attenuation with the IR to UV
flux ratio greater than 10 and they are often  not detected in UV (cf
Fig.~\ref{lIR_UV-lbol}). It may be noted that the fraction of non
detections does not vary a lot with the bolometric luminosity of the galaxies. The
IR selection does not include the large number of very blue galaxies
detected in the UV with  low bolometric luminosity (cf
Fig.~\ref{lIR_UV-lbol}). Last of all, when only galaxies selected at both
wavelengths are concerned (dashed histograms) the properties of 
both samples are found to be similar, it is also clearly seen in Fig.~\ref{lIR_UV-lbol}.

\section{Specific star formation rates}
\subsection{ Variation in the SSFR within each sample}
\begin{figure*}
   \centering
   \includegraphics[angle=-90,width=15cm]{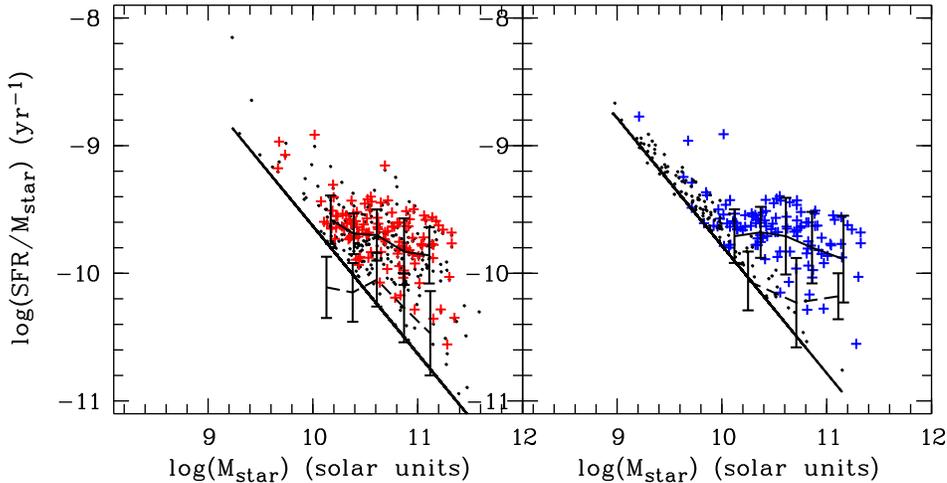}
      \caption{SSFR versus stellar masses. Left panel: IR selection, the z=0.7 galaxies detected at 24 $\mu$m and in UV are plotted with red crosses, the galaxies only detected at 24 $\mu$m with dots. The solid line with error bars (1 $\sigma$) is the result of the volume average at z=0.7, the dashed line the volume average at z=0. The diagonal line represents the detection limit at z=0.7. Right panel: Same plot for the UV selection. At z=0.7 galaxies detected at 24 $\mu$m and in UV are plotted with blue crosses, the galaxies only detected at UV with black crosses, the lines are defined in the same way as for the IR selection. Adopting $\eta$=0 instead of $\eta$=0 at z=0.7 would shift all the data at z=0.7 by -0.15 dex along the vertical axis }
              
         \label{ssfrobs}
\end{figure*}

In this section, we analyze the variations in  the specific star
formation rates (SSFR, defined as the total SFR divided by the stellar mass)
as a function of the stellar mass within our samples of star-forming galaxies.
In Fig.~\ref{ssfrobs}  the SSFRs versus stellar masses are plotted for
both samples at both redshifts. Average trends are
plotted at z=0. They come from \citet{buat06} where volume averages were
performed. The values were  converted  using the 
SFR and stellar mass calibrations adopted in this paper (section 2.1). At
z=0.7, the values for each galaxy are plotted.

Before any interpretation, we must estimate the detection limits in
SSFR to determine where the results are reliable, on the basis
of the limiting fluxes adopted in UV and IR.
Stellar masses were estimated from the 3.6 $\mu$m
flux, available for each galaxy, and are thus unaffected by
detection limits.
Practically speaking, for the IR selection we adopt the limit of $\log(L_{\rm IR}) =
10.34~(\rm L_{\odot})$ used to build the sample (cf section 2.2.1) and translated it into 
SFR at z=0.7 with Eq. 1. We obtain $SFR_{\rm lim}=2.34 ~\rm M_{\odot} yr^{-1}$.
Given the  high  values of the IR to UV ratio for the IR selection (cf
Fig.~\ref{lIR_UV-lbol}), we have neglected the contribution of the UV
emission to estimating $SFR_{\rm lim}$.  For the UV selection, the adopted
limit $\log(L_{\rm UV})=9.39 (\rm L_{\odot})$ translates into an ``unobscured'' SFR at z=0.7
of 0.50 $\rm M_{\odot} yr^{-1}$ according to Eq. 2. However we must also
account for the contribution of the IR emission to the SFR which is
not negligible at all (cf Fig.~\ref{lIR_UV-lbol}). We used the
detection limit of 0.016 mJy at 24 $\mu$m (cf. section 2.2.1)
translated into an ``obscured'' SFR at z=0.7 of 1.15 $\rm M_{\odot} yr^{-1}$. The  limit thus obtained for the  total
SFR is $SFR_{\rm lim}= 1.65 \rm M_{\odot} yr^{-1}$. The resulting detection
limits obtained for z=0.7 for the SSFRs (and only constrained by those
on the SFR) are reported in Fig.~\ref{ssfrobs}.

The z=0 samples are purely flux-limited so we cannot estimate the detection limits in the same way as for the z=0.7 sample. \citet{buat06} built bolometric luminosity functions  with these samples ($L_{\rm bol} = L_{\rm UV}+(1-\eta)~L_{\rm IR}$) down to $\rm 6.3~10^8~L_{\odot}$ and $\rm 1.8~10^9~L_{\odot}$ for the UV and IR selections  respectively. These luminosities  were used to estimate the limits reached in SFR and  they can be translated in SFR using Eq. 1. We obtain  $SFR>0.07 \rm M_{\odot} yr^{-1}$  for the UV selection and $SFR>0.19 \rm M_{\odot} yr^{-1}$ for the IR selection. Such low SFRs put detection limits well below the values reported in Fig.~\ref{ssfrobs} for the mean trends at z=0.

The influence of the choice of $\eta$ (Eq. 3) can be easily checked:
the measure of the SFR is mostly dominated by the IR emission with
only a small contribution from the UV one, especially for massive
galaxies which will be the main topic  below.
Therefore adopting $\eta=0.3$ instead of $\eta=0$ at z=0.7 will reduce
the SFR and hence the SSFR by a factor 0.7, i.e. a translation of -0.15 dex
on the vertical axis of Fig.~\ref{ssfrobs}.

It is obvious from Fig.~\ref{ssfrobs} that the locus of the galaxies
with masses lower than $\rm 10^{10}~M_{\odot}$ is completely governed by
detection limits at z=0.7. The observed trends are only reliable for
$M>10^{10}~\rm M_{\odot}$. In the following, we limit our analysis
to this mass range. The trends found for both selections seem to be
similar, they will be compared in the next sections.

Our samples were built to be complete in terms of the star formation
  rate (as discussed above in this section) but are not expected to be
  complete in mass. Indeed a mass-selected sample also contains
  quiescent galaxies that are not present in our sample. At z=0 we built the stellar mass function using the V/V$_{max}$ formalism for the IR and the UV-selected samples. They are compared to the stellar mass functions derived by \citet{belletal03} from SDSS+2MASS data. We have applied the correction preconised by \citet{belletal03} for a Kroupa IMF. As expected, the stellar mass distribution of the UV-selected sample is similar to what is found by \citet{belletal03} for late-type galaxies (selected according to their colors), whereas the IR selection leads to a larger number of massive galaxies but still below the total stellar mass distribution. 
 At z=0.7 we can directly
  compare the counts obtained for a selection at 3.6 microns (IRAC),
  24 microns (MIPS), and in NUV (GALEX) for the redshift bin 0.6-0.8.
   Fig.~\ref{mstar} summarizes the comparison where the counts are
  reported against the stellar mass. The selection at 3.6 microns
  consists of  sources brighter than 3 $\mu$Jy, which corresponds to
  $\rm 10^{10}~M_{\odot}$ at z=0.7; this limit is very conservative
  given the depth of the IRAC observations
  \citep{elbaz07,sanders07}.  Both MIPS and GALEX selections lead to
  lower counts than the IRAC one. The effect is more extreme for the
  UV selection. This was expected from Fig.\ref{lIR_UV-lbol} where it is
  obvious that the UV selection misses bright (and therefore massive)
  galaxies with a substantial dust attenuation. Conversely the
  contribution of the UV sources becomes significant for masses around
  $\rm 10^{10}~M_{\odot}$. We have also included in Fig.~\ref{mstar} galaxies selected in
  either the UV or the IR in order
  to analyze their total mass distribution. The
  difference between the number of galaxies in our GALEX+MIPS
  selections, and the IRAC one is likely to be due to the presence of
  quiescent systems detected neither at 24 $\mu$m nor in NUV.  It is
  qualitatively consistent with the relative contribution of spirals
  and irregulars to the total stellar mass function as measured by
  \citet{bundy05} in the same redshift range. A full analyzis of the
  luminosity functions in our selected samples is in preparation
  (Takeuchi et al.).

\begin{figure}
   \centering
   \includegraphics[angle=-90,width=13cm]{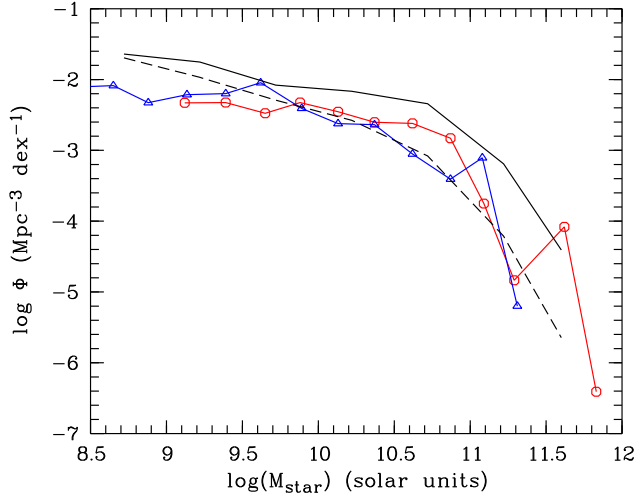}
      \caption{Stellar mass functions for the z=0 samples selected in IR (circles) and in UV (triangles). The solid line without symbols is the total stellar mass function derived by \cite{belletal03} and the dashed one denotes the stellar mass function for late type galaxies by the same authors}
              \label{mstarz0}
\end{figure}

\begin{figure}
   \centering
   \includegraphics[angle=-90,width=13cm]{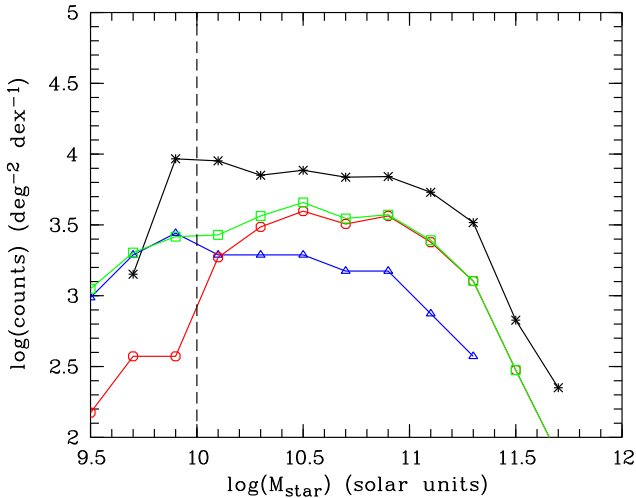}
      \caption{Galaxy counts as a function of the stellar mass for a selection at 3.6 $\mu$m (stars), 24 $\mu$m (circles), in NUV (triangles) and for the galaxies selected either at  24 $\mu$m or in NUV (squares)}
              \label{mstar}
\end{figure}

Before comparing to models we  also calculate average trends at
 z=0.7 (Fig.~\ref{ssfrobs}). There is no need
for volume corrections (cf section 2.2) and simple mean values can be 
calculated. We have used the Kaplan-Meier estimator for the IR selection and the error bars plotted in Fig.~\ref{ssfrobs} correspond to the   1$\sigma$ dispersion. For the UV selection, the presence of upper limits located on one side of the distribution hampers the use of the Kaplan-Meier estimator. Therefore we have calculated the median instead of the mean for the UV selection at z=0.7. Very similar trends are found in both
selections (see also Fig.~\ref{IRfit} where average trends for both
selections are overplotted). Even if the UV selection misses 
highly reddened galaxies (cf Fig. ~\ref{lIR_UV-lbol}), the star formation
activity of the galaxies selected in UV is found to be similar to those
selected in IR: the SSFRs do not depend critically on the adopted selection (as long as they are estimated by combining UV and IR emissions), a result already found at z=0 once volume corrections are applied to the local samples \citep[][and Fig.~\ref{IRfit}]{buat06}. 

\subsection{Comparison with previous studies at intermediate redshift}

Numerous works address the question of the variation in the
specific star formation rate at various z. Whereas these studies
conclude to a general decrease of the SSFR with galaxy mass even at
high z (the so-called downsizing effect), the trends exhibit large
departures from  one work to another. Our aim in this section is not to perform an
exhaustive comparison with all the available results: we will instead discuss
a few studies performed in the same redshift range as the present work,
and using methods to estimate SFR and stellar masses similar to ours.

\citet{zheng07a} analyzed the dependence of star formation on galaxy
mass within the COMBO 17 survey adding SPITZER data to measure
accurate SFRs. The SSFRs they reported are consistent with ours within
error bars but with a steeper decrease in the SSFR when stellar mass
increases  compared to  what we find. Their sample contains blue and red
galaxies, \citet{bell07} considered them separately and found a
flatter distribution of SSFR for the blue ones than for the red ones.
In a similar way \citet{elbaz07} analyzed the GOODS survey and found a
flat distribution of SSFRs at z=0.8-1.2 and between $10^{10}$ and
$\rm 10^{11} M_{\odot}$. The decrease in the SSFR they found around $\rm 10^{11}
M_{\odot}$ is clearly due to the red galaxies, whereas blue galaxies
exhibit a roughly constant SSFR (their figure 17).

\citet{noeske07} also found a flat distribution that is fully consistent with
ours within the AEGIS survey (both in absolute values and trends) for
galaxies with stellar masses  between $10^{10}$ and $\rm 10^{11}
M_{\odot}$ and  more massive galaxies have a lower SSFR. Studies based on
GALEX data and combining UV and IR emissions have also led to an
almost flat distribution of SSFRs. \citet{iglesias07} performed a UV
selection combined with SWIRE data from z=0.2 to z=0.7 and found a
flattening of the SSFR versus M$_{\rm star}$ variation from z=0.2 to
z=0.7. \citet{martin} also found a flat distribution in the CDFS at
intermediate z with deep GALEX and MIPS data. \citet{zamojski} perform
a very complete analysis of the COSMOS field (although only based on
UV-optical rest-frame data) at z=0.7. Their UV-detected sample also
exhibits a rather flat distribution of SSFR, much flatter than  
obtained for their entire sample (including objects not seen in UV).

Although a large dispersion is found in the already published results,
we find  good agreement with those  focusing on star-forming
galaxies alone. The steeper decrease in the SSFR with increasing mass
found in some studies at intermediate redshift is likely to be due to the presence of quiescent
systems, which are absent in our present selection
\citep[e.g.][]{bell07,elbaz07}. Similar  differences are found at z=0 when only star-forming galaxies are selected or quiescent are also added \citep{brinchmann04,buat06,elbaz07}. Most of the decrease found at z=0.7 also occurs for
masses higher than $\rm 10^{11} M_{\odot}$
\citep[e.g.][]{zheng07a,noeske07,elbaz07} which are not well
represented in our sample.\\

\section{The models}
\begin{table*}%t2
\caption{Models with $\lambda=0.05$, $z_f=6$, stellar masses are calculated at z=0 (column 2) and z=0.7 (column 3)  with the analytical fit}\label{table2}
\centering
\begin{tabular}{cccccc}
\hline \hline
 velocity (km s$^{-1}$) & $\log(M_{\rm star})_{z=0}(\rm M_{\odot})$)&$\log(M_{\rm star})_{z=0.7}(\rm M_{\odot})$ & a & b & c  \\
\hline
80 &8.89&8.19&6.62&0.41&0.36\\
150&9.92&9.51&8.74&0.98&-0.20\\
220&10.52&10.25&10.01&1.25&-0.55\\
290&10.94&10.75&10.81&1.35&-0.74\\
360&11.25&11.10&11.35&1.37&-0.85\\
\hline
\end{tabular}
\end{table*}
 
Recent studies have found evidence of the minor role of
strong mergers in the evolution of massive galaxies from z=0 to
$\sim$ 1 \citep{bell05,melbourne,zheng07b}. Therefore it is tempting
to try to fit the data with models that assume a smooth average evolution, which is
weakly challenged by small interactions and minor mergers.

Recently \citet{noeske07} proposed a simple model of gas exhaustion to
interpret the variation in the SFR as a function of the stellar mass
out to z=1.1. In their model galaxies experiment exponential star
formation histories with e-folding rates and redshift formation
varying with the galaxy mass, less massive galaxies being younger and
having higher e-folding rates.

Here, instead of building ad hoc models \citep[the parameters in ][are
fine-tuned for their data]{noeske07} for our study, we follow a
``backward'' approach \citep[e.g.][]{silk99} by starting from studies
of the Milky way and the nearby universe to extrapolate the behavior
of galaxies at higher z, without any further adjustment of the models that are constrained in the local universe (see details below).

\subsection{ Description of the models}

To interpret our results with physically motivated models, we
 use a grid of models for the evolution of spiral galaxies
 that is similar to the one presented in Boissier
\& Prantzos (2000). These models were calibrated to reproduce many
properties of the Milky Way (Boissier \& Prantzos, 1999), and were compared
to nearby spirals in subsequent works. They should be adequate to broadly represent
 the family of spirals from very massive ones to irregular low-mass
exponential disks.
The models use scaling relationships to simulate disks of various
rotational velocity and spin parameter \citep[$\lambda$, measuring the
specific angular momentum, see e.g.][for a definition]{mo98}.  The velocity is closely related to the
total mass of the galaxy and, as a result, its final stellar mass.
Each stellar mass (at any redshift) corresponds to one and
only one model for a given specific angular momentum $\lambda$.
The specific angular momentum has a log normal distribution
\citep[e.g.][]{mo98}.  Spirals have spin parameters between
0.02 and 0.08, with a typical value of 0.05 \citep[the Milky Way Galaxy
corresponding to $\lambda \sim$ 0.03 in][]{boissier00}.
Since we are interested in general trends with mass, we only use 
models with this average $\lambda=$0.05 in the following. The quantities we
present in this paper present, in any case, a much stronger
dependence on the velocity than on the spin parameter.

The models assume that galaxies are formed by progressive infall of
primordial gas, starting at high redshift ($z_f = $ 6).  Models of the
Milky Way shown early-on that infall of ``fresh'' gas provides a good
explanation especially for the distribution of metallicity of G-Dwarfs
in the Solar neighborhood and the idea has been widely used
\citep[e.g.][]{larson72,pagel97,chiappini97,boissier99}.  Such models do
not state that this gas has been forever in a reservoir around the
galaxy. In a more modern context, it could very well be
that this infall corresponds to small satellites (with large gas
fraction) being accreted by the galaxy during minor mergers.
Inside the disk, stars form from the gas following a Schmidt-like law,
including a dynamical factor.  In a recent work \citep{boissier03},
the star formation law was empirically determined to be 
\begin{equation}
\Sigma_{\rm SFR} =  2.63 \, 10^{-3} \Sigma_{\rm GAS}^{1.48} V(R)/R
\end{equation}
relating the surface density of star formation rate $\Sigma_{\rm SFR}$
($\rm M_{\odot} pc^{-2} Gyr^{-1}$) to the gas surface densities
$\Sigma_{\rm GAS}$ ($\rm M_{\odot} pc^{-2}$). Here $V(R)$ is the rotation velocity
($\rm km \, s^{-1}$) at radius $R$ (kpc). This formulation is very close to the
one originally used, and subsequent models (including in this paper)
 have used it. The results are very close to those in
\citet{boissier00}, and show the same global trends.

The star formation history in these models  strongly depends  on the
built-in assumption that infall proceeds at a higher pace early in the history
of massive galaxies with respect to the Milky Way, and later in lower mass
galaxies. This assumption corresponds to the so-called ``downsizing'' 
effect, and was made to adequately reproduce observed trends in nearby spirals.

By construction, these models should  match the z=0 universe at
the current epoch.  
The use of such models in a backward approach is indeed justified
only because a very large number of properties of nearby galaxies
(i.e. the redshift zero universe) are correctly reproduced by them.
It is notably the case (and we refer to the references for further
details) of scaling relationships such as surface brightness and
scalelength versus magnitude, the Tully-Fisher relationship (and its
dependence on wavelength), color-magnitude diagrams (B-K vs K),
luminosity-metallicity relationship, spectra \citep{boissier00},
colors and abundance gradients in spirals \citep{prantzos00}, star
formation rates and  gas fractions \citep{bobo01}.

In a ``backward'' approach, their prediction about  the past was 
only compared to a few properties of small samples of spirals at
higher redshift \citep{boissier01}, and no serious discrepancies were
found with the data available at the time.   Note that the
  observed B-band luminosity function at redshift zero was reproduced
  by the models in this paper. This result was obtained by
  construction since the authors assumed a circular velocity
  distribution $V_C$ derived from observations, and the models reproduced the $M_B$ vs $V_C$
  relationship.

We propose here to use these models to interpret  
data concerning the SSFR (such data did not exist a few years ago).
\begin{figure*}
   \centering
   \includegraphics[width=7cm]{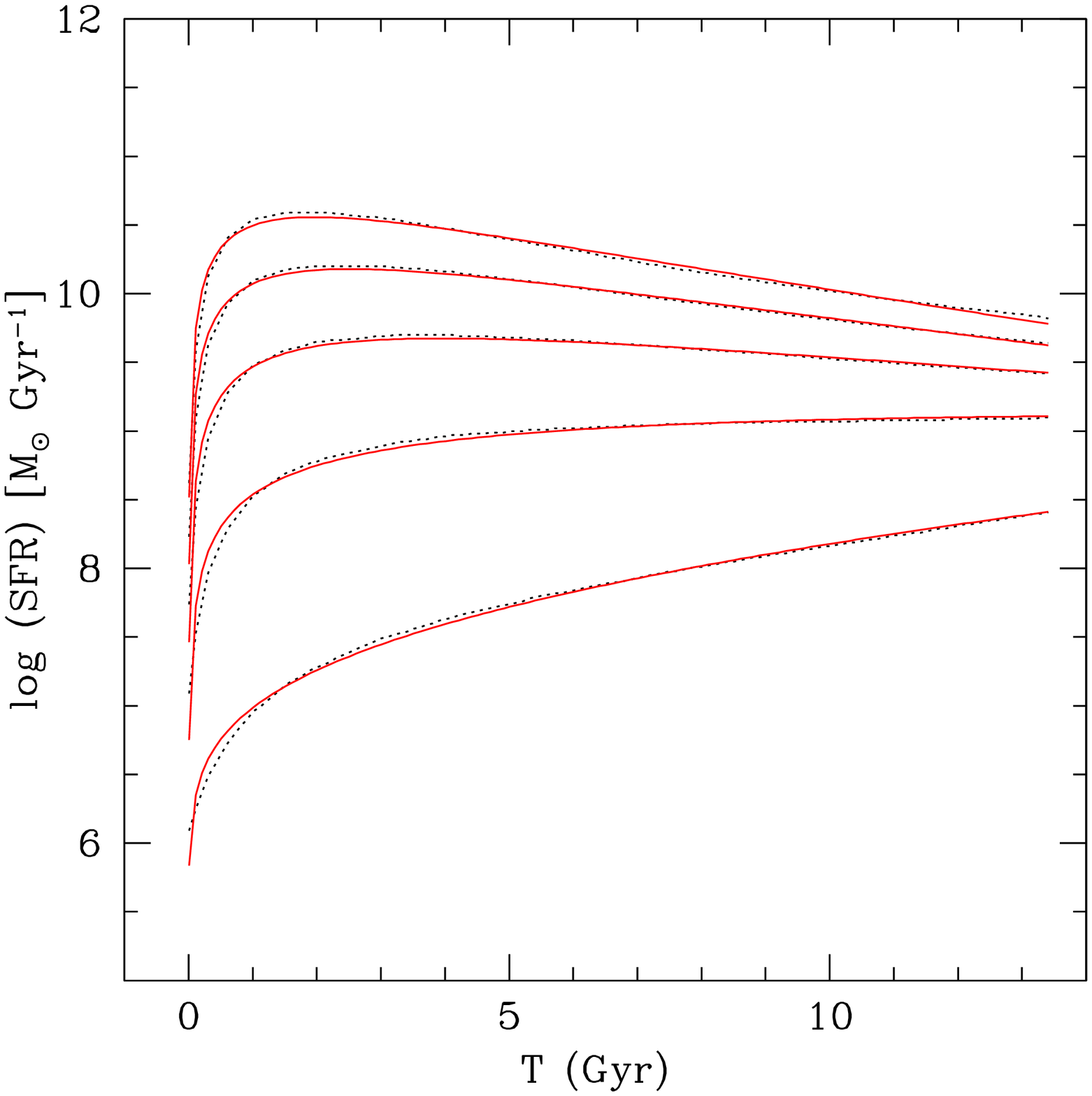}
   \includegraphics[width=7cm]{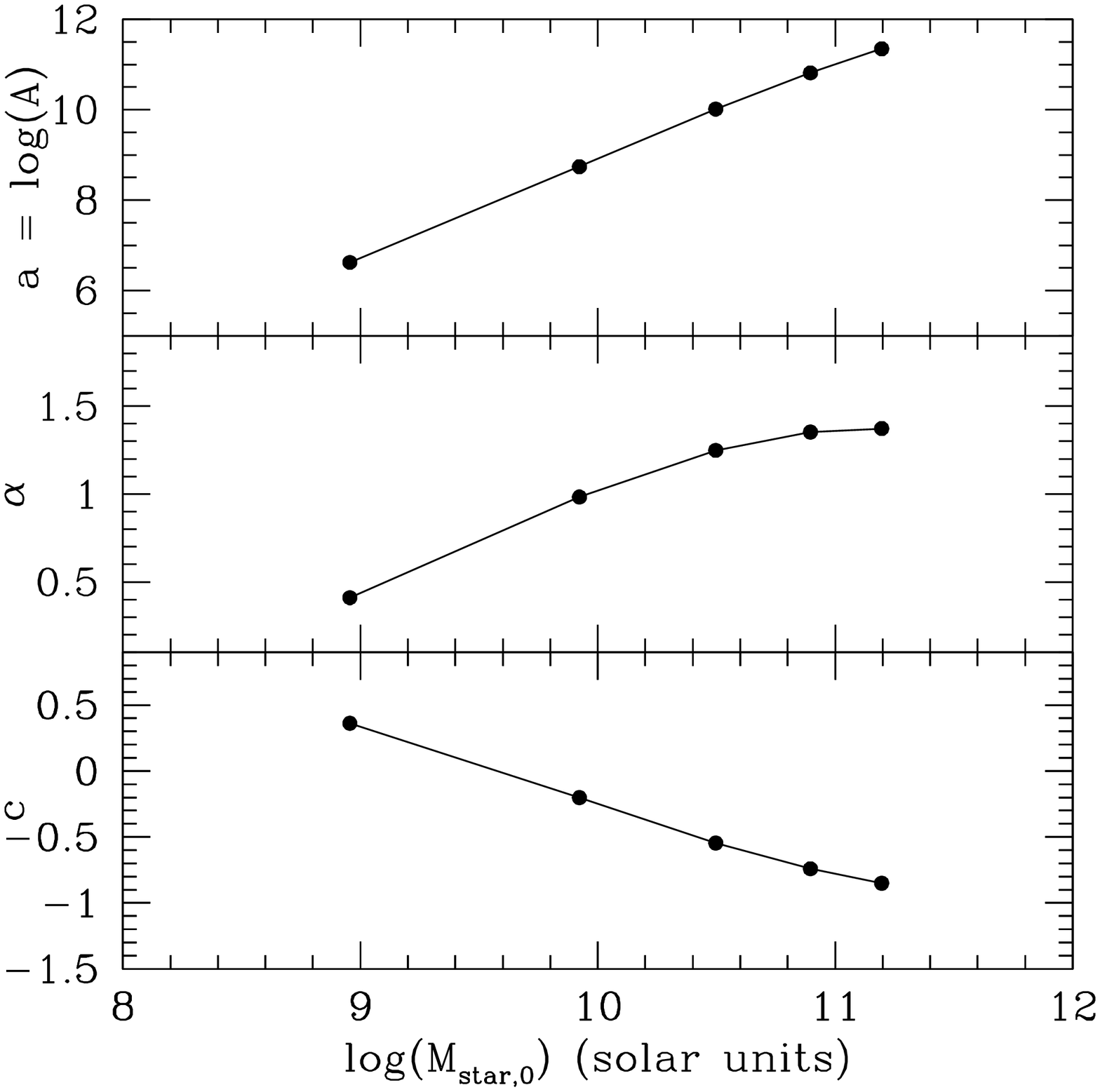}
   \caption{Left: Star formation histories of the adopted models for
     five final stellar masses (equivalently: five rotational velocities), 
     from the lowest mass (bottom) to the highest one (top). 
     The dotted line is the result
     of  computing of the chemical evolution models and 
     the solid line is the analytical fit of (Eq. 6). On the x-axis, T=0
     Gyr corresponds to the formation of the galaxies ($z_f=6$), z=0.7
     corresponds to T = 6.2 Gyr and z=0 to T = 12.5 Gyr. Right:
     Variation in  the three parameters (a,b,c) of the analytical star
     formation history (Eq. 6) as a function of the present (z=0)
     stellar mass (values are given in Table 2 for the five models shown on the left, also
     shown as dots in this panel).}
    \label{sfrmod}
    \label{parammod}
   \end{figure*}

\subsection{ Star formation histories}

The star formation histories (presented in Fig.~\ref{sfrmod}, left panel) of the
models  result from the non linear combination of  a star
formation law, an infall history and its mass dependence. To
provide an easy way to compare our star formation histories to others,
or to various data sets, we propose  an analytical form
for the star formation history of our models. After testing several
possibilities, we obtained  a correct fit using three parameters:

\begin{eqnarray}
 SFR(t) = A t^{b} \exp(-(t/\tau)^{0.5})~~ {\rm or}\\
\log(SFR(t)) = a+b~\log(t)+c~t^{0.5 \label{fitsfrh} }\\
{\rm with} ~~a = \log A~~~ c=-0.43 \tau^{-0.5}.
\end{eqnarray}

We can assign a tentative physical significance to each of the
parameters : $A$ is a scaling factor, $b$ is measuring the rate at
which the star formation increases early-on, and $\tau$ is
a time-scale for the decay of star formation at later time
(this is not always  the case: for low-mass galaxies
the star formation rate is still rising at late epochs).
The values of these parameters for the $\lambda=0.05$ models are given
in Table 2 for a few values of the circular velocity and their dependence on the final (z=0)
stellar mass is shown in Fig.~\ref{parammod} (right panel). The
quality of the fit can be judged from the left panel.
With such a formula, it is straightforward to compute analytically the
evolution of stellar masses assuming the instantaneous recycling
approximation \citep[e.g.][]{pagel97} and a returned fraction of R=0.3 for the 
\citet{kroupa93} IMF \citep{boissier99}. We checked that this  simple 
approach gives a very good approximation  to the results of the numerical
computations implementing infall, star formation law, and finite lifetimes of stars.

Before comparing the models and the data, we should note that some
ingredients of the chemical evolution models are rather uncertain.
Especially, the star formation law efficiency is very dispersed among
disk galaxies, making the star formation rate uncertain by about a
factor 2. In addition,  we used the
\citet{kroupa93} IMF, for consistency with earlier works.  We checked that if we used the more recent
\citet{kroupa01} IMF for the models, the star formation rate and
stellar mass during the overall history of the galaxy would change by
only 10 to 20 \%.  Modifying the IMF from \citet{kroupa93} to
\citet{kroupa01} mainly affects the population of massive stars and,
as a consequence, the calibration of the current star formation rate
and the predicted UV fluxes is  why we use the more
recent IMF \citep{kroupa01} to derive SFR from the UV fluxes (Eq. 2).
Model predictions other than UV fluxes depend much more weakly
on this choice.

%%%%%%%%%%%% ON THE HAMMER & CO %%%%%%%%%%%%%%%%%
\citet{hammer07} suggest that the Milky Way,
having escaped significant merging over the last $\sim$ 10 Gyr,
is in fact unrepresentative of spirals in contrary to M31.
Our models, even if they were calibrated in the Milky Way,
successfully reproduced many properties of
nearby spirals, including abundance and colour gradients
\citep{boissier00,prantzos00}. 
%%%
The star formation law (the ``Schmidt''-like law mentioned above) was
determined from 16 galaxies with H$\alpha$ profiles
\citep{boissier03}, and is consistent with the SFR indicators in the
Milky Way. The SFR profiles have been determined in 43 nearby
spirals from UV data \citep{boissier06} and compared to gas profiles.
The SFR-gas connection they find agrees with the one derived from
H$\alpha$ profiles, even if the relationship extends to lower gas surface densities
in the UV (the untypical galaxy with this respect was actually M31,
which does not follow the Schmidt law in its inner part).
We recognize that the models calibrated in a relatively calm Milky Way
 probably not include all the details of disk galaxy evolution.
However, they do  agree with many properties in nearby spirals
as discussed above and we believe they reproduce the major traits of
disk evolution even if real galaxies are likely to sometimes suffer interactions not taken into account in this approach (as long as
they are not dramatic events destroying the disks such as  major
mergers.

\section{Results and discussion}

\subsection{ Comparison between models and data at z=0 and z=0.7} 

\begin{figure}
  \centering
  \includegraphics[angle=-90,width=13cm]{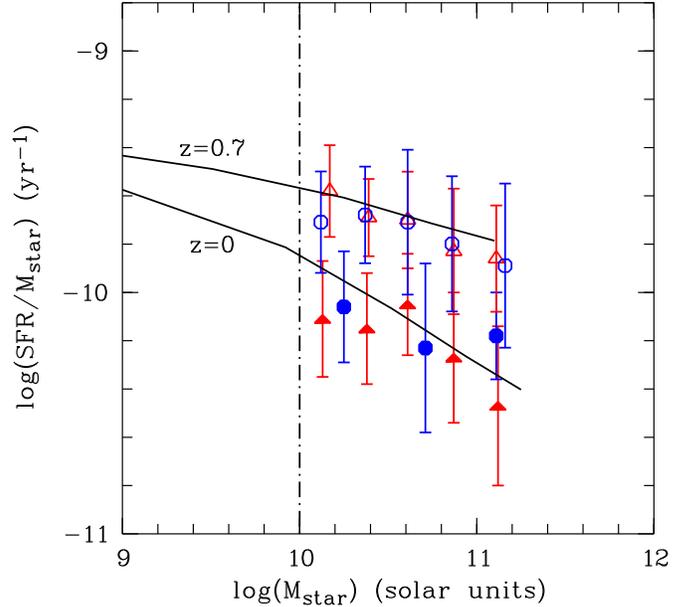}
  \caption{Specific SFRs versus the stellar mass: comparison between data and model. UV selection: blue  circles and error. IR selection:red triangles and error bars. The filled symbols  are for z=0, the empty ones for z=0.7. The solid lines represent the model summarized in Table 1 (see text)}
  \label{IRfit}
\end{figure}

In Fig.~\ref{IRfit}, the results of the model are compared to the
mean observational trends obtained in section 4. The agreement can be
considered as good for both selections at $M>10^{10}~\rm M_{\odot}$
given the large intrinsic dispersion of the data and the fact that no
fitting has been performed. 
At z=0.7 the agreement is remarkably
good given the uncertainties inherent to both the models and the data
(see the discussion below).  At z=0 the SSFRs predicted by the model
lie slightly above most of the mean observed values and exhibit a
steeper decrease as the stellar mass increases, as compared to the UV
selection, although the intrinsic dispersion of the observed
quantities is very large.  As discussed in section 5.2, models
themselves contain some uncertainties. Modifying the star formation
law efficiency within the uncertainties could modify the star
formation rate by a factor $\sim$2, and various IMF may modify the
results of the models by up to $\sim$ 20 \% as seen above. The models errors are likely to be systematic if the IMF and the
star formation are universal, i.e.  directly linked to the physics of
star formation on a local scale rather than depending on the global
properties of galaxies. In that case, the trends obtained by the
models are robust even if the absolute values could change. \\

Models predict  directly the stellar masses, their
estimates do not depend on any assumed mass-to-light ratio.  The
uncertainty on the mass-to-light ratios only affects the masses
derived from the observations. It can reach a factor 2
\citep{kanna07}, similar to the dispersion found for the observed
SSFRs for a given range of stellar mass.
Therefore both models and
observables are likely to be affected by systematic errors large
enough to explain the slight shift found at z=0 between the models and
the data in Fig.~\ref{IRfit}. Another issue may be that the SSFR
trend with the
stellar mass is flatter for the UV-selected sample than the models
predictions. Indeed the UV selection leads to a selection of late-type
systems (cf section 4.1 and Fig.~\ref{mstarz0}); and since the
contribution of early type galaxies increases with the stellar mass
(Fig.~\ref{mstarz0}), we expect a flatter distribution of the SSFR for
the late type systems alone as compared to the total galaxy
population. The models are set up to reproduce the mean evolution of
galactic disks.
%which also involves relatively quiescent disks at z=0, 
For massive galaxies, they correspond to relatively early disk
types (as early as Sa) in which the specific star formation rate is low: i.e.,
the star formation activity is low.

The UV selection can pick up massive galaxies with active star
formation due to burst or refueling through interactions (leading to
later type galaxies for the same mass) not introduced in the models
that  then correspond better to the galaxies from the IR selection (picking
up massive galaxies with more quiescent star formation).

At higher z, because of the
higher star formation activity in all the galactic disks, this effect
is expected to be smaller and the discrepancy between the models and
the data disappears.

\citet{noeske07} have performed a similar analysis to ours. They selected
galaxies between z=0 to z=1 from the AEGIS survey. They derived SFR by
combining IR, optical, and UV data. Their SSFR distributions
are very consistent with ours. As a result, the gas exhaustion 
model they propose would also fit  our data reasonably well.
In their model, star formation rates follow an exponential 
decay after a formation redshift $z_f$.
Less massive galaxies have a longer e-folding rate and a lower
redshift of formation. Nevertheless, it implies a wide range of
redshift formation, from $z_f$=1 for $\rm 10^{10} M_{\odot}$ up to $z_f$=3
or $\rm 10^{11} M_{\odot}$. In our model, on the contrary, all the
galaxies have the same formation redshift $z_f=6$, which is the epoch
when galactic building blocks are assumed to start to exist.  It is
roughly the redshift when the first galaxies are confirmed to exist
\citep[e.g.][]{schaerer07}. In this approach, it is the time variation
of the SFR that depends on the galaxy mass, in such a way that the
bulk of star formation occurs at different ages according to the
galaxy mass.
Actually, the star formation histories obtained in our models are
qualitatively similar to the schematics proposed by \citet{sandage},
which inspired more recent works.  Among them, \citet{gavazzi02} used
a star formation history law ``a la Sandage'', which is mimicking the
\citet{sandage} trends. Quantitatively, our star formation history is 
slightly different: for massive galaxies especially, the SFR does not
decrease as quickly at late epochs.

Because of this, the mean evolution of the SSFR from z=0 to z=0.7 in star-forming galaxies of $\rm 10^{10}-10^{11} M_{\odot}$ is consistent with the
one predicted by a simple but physically motivated model of secular
evolution, in which the galaxies are progressively built by accretion
of low-metallicity gas  and in which star forms according to a
``universal'' law of star formation. It does not mean that individual
galaxies cannot experience any burst events or interactions, but our aim
is only to reproduce mean trends. At z=0, active galaxies were found in
the IR selection \citep{buat06}, but these objects were also diluted
among the more numerous fainter objects when volume averages were
performed. In the same way the most active galaxies of our samples at
z=0.7 have SFRs reaching 40 $\rm M_{\odot}~yr^{-1}$, a factor $\sim 2$
larger than highest SFR from our models (Fig.~\ref{parammod}). Indeed
luminous infrared galaxies (LIRGs which account for 30 $\%$ of our IR
selection) are known to experience bursts \citep[e.g.][]{marcillac}.
Nevertheless when averages are calculated their influence is diluted.
Indeed, our results do not imply a secular evolution for the whole
galaxy population but are instead consistent with a smooth mean 
evolution in agreement with the idea that galaxies are built 
by infall (which can be in the form of minor mergers, as suggested above). 
In this context, major mergers or other extreme events are not expected to 
participate much in  the construction of the average galaxy.

Interestingly, the model with a circular velocity of 220 km/s,
corresponding to the one of the Milky Way, has a stellar mass of $\log
(M_{\rm star})~=~10.52~ (\rm {M_{\odot}})$ at z=0 and $\log(M_{\rm star})~=~10.25~ (\rm{ M_{\odot}})$ at
z=0.7. In other words, our samples probe Milky Way analogues in terms
of stellar mass.

 \subsection{Comparison with semi-analytical models}

The models presented above have the advantage of reproducing many properties of
galaxies at redshift zero.
%(and not just selected ones, such as e.g. the Tully-Fisher relationship). 
They assume smooth evolutions on average, which  might not be 
realistic in the  paradigm of hierarchical galaxy formation.
Semi-analytical models (SAMs) usually follow a similar approach to our models
(e.g. assuming empirical recipes for star formation laws and scaling relationships)
but obtain the mass accretion and merging histories of the galaxies by following the 
hierarchical growth of dark haloes (and the baryonic galaxies within them). 
%
%As such, they should be more realistic than our simple models and 
%correctly reproduce e.g. the properties of galaxies
%suffering massive merger events.

\begin{figure}
  \centering
  \includegraphics[angle=-90,width=13cm]{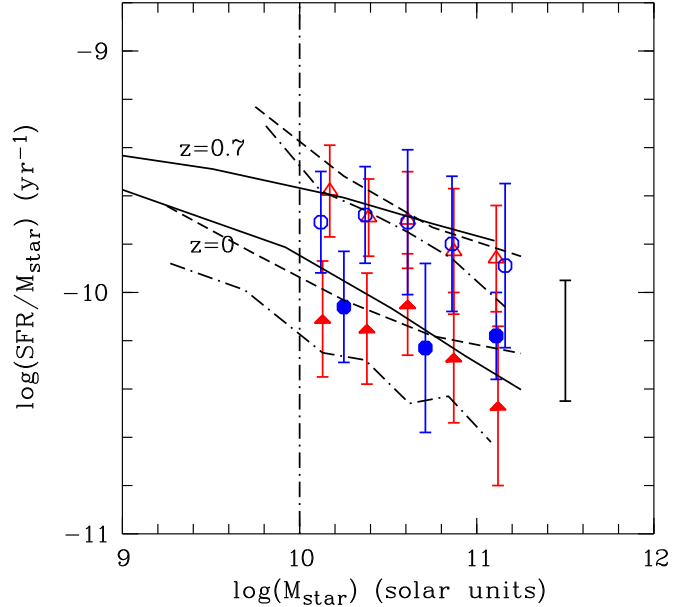}
  \caption{Predictions of the Millennium model at z=0.7  and z=0 (dot-dashed lines) and of the numerical catalog of \citet{nagashima05} (dashed lines). All the other lines and symbols are the same as in Fig~\ref{IRfit}. The mean error bar (1$\sigma$) is indicated on the right side of the panel}
  \label{SAM}
\end{figure}

We also compared
our data to the results of two sets of SAMs. \citet{kitzbichler07} have produced
lightcones simulations 
that can be very conveniently compared
to observations. They
applied the SAM of \citet{croton06} to the Millennium Run simulation
\citep{springel05}. \citet{nagashima05} have constructed a numerical
catalog also based on SAMs and combined with high resolution N-body
simulations. We have applied selections to these simulated catalogs as
close as possible to those obtained for our observational data
(section 4.1): an SFR higher than 2 $\rm M_{\odot}~yr^{-1}$ for
galaxies at a redshift  between z=0.6 and 0.8 and an SFR
 higher than 0.1 $\rm M_{\odot}~yr^{-1}$ for nearby galaxies ($z<0.1$).
We computed average trends with the simulated data in the same way as
in section 4 for our observational datasets.  The results are
displayed in Fig.~\ref{SAM}, where  the simulated quantities are compared to our
mean observational trends obtained for the IR and UV selected samples.
It can be seen that the simulated data also reproduce  the values
of the mean SSFR observed at z=0.7, although the simulations lead to a
somewhat steeper decrease in the mean SSFR towards high stellar masses
than observed. A rather large discrepancy is observed at z=0 for the
millennium simulations: the simulated galaxies have lower SSFRs than
the values measured in our samples. If we ignore this discrepancy the
agreement is satisfactory. We can conclude that SAMs 
present a similar mean behavior to the one found 
for our simpler evolution model
aimed at reproducing present day spirals without strong merger events.
%(expected since the physical recipes are the same...??Samuel??). 
This result reconciles backward and forward approaches for the evolution of
galaxies at least until z=0.7 and validates our simple approach for
star-forming galaxies. \\

The SAMs should be able to reproduce galaxies with very high SSFR induced by major mergers, whereas our model only aims at reproducing mean trends. 
Nevertheless, it may be meaningful to mention that the timestep used in 
the simulation (about 300 Myr for the Millennium simulation 
\citep{kitzbichler07}) might be too large to catch short time bursting 
systems and the SFRs averaged over such a time step cannot be very high. 
This aspect of the simulation works should be improved in order to make 
a more extensive comparison of star-formation related quantities, like SSFR.
A more exhaustive discussion of the results of semi-analytical simulations 
is beyond the scope of this paper and will be shown elsewhere.

\subsection{Predictions and comparison with data for higher z}

Since our models are able to reproduce the evolution of massive star-forming galaxies from z=0 to z=0.7, we can take a step further and make some 
predictions at higher z. 
These predictions can be considered as a reference for quantifying the
expected properties of distant galaxies in the absence of major
merging, implying a strong starburst and a very high  efficiency of  star formation. 
In Fig.~\ref{predicthiz} are reported the evolution up to z=4 of the
SSFR and of the gaseous phase metallicity of the chemical evolution
models.

\subsubsection{Redshift evolution of the specific star formation rate}

\begin{figure*}
  \centering
  \includegraphics[width=15cm]{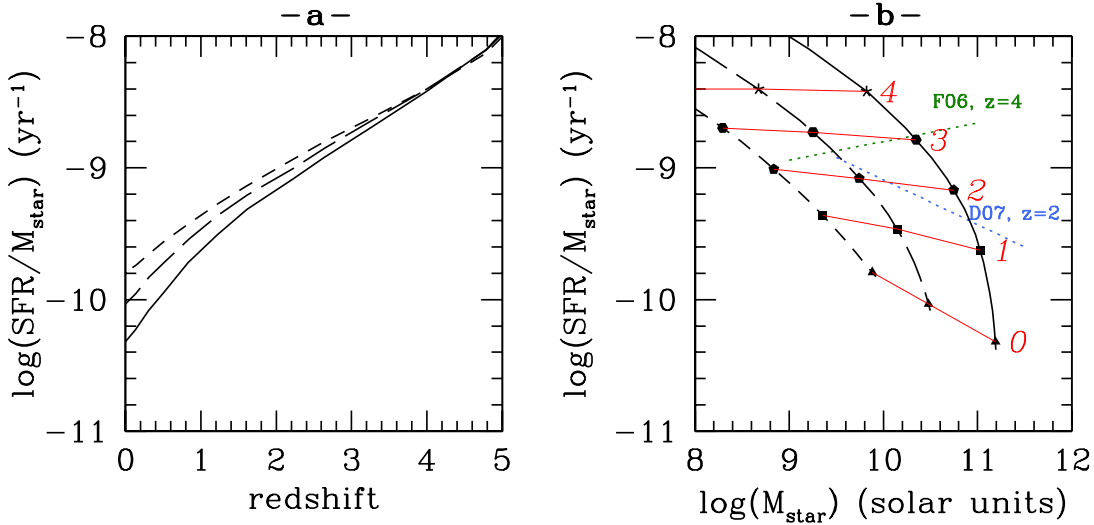}
  \caption{Predictions of the models for higher z. All quantities are
    in solar units. Panel a: SSFR as a function of z for three
    rotational velocities and therefore masses (cf Table 1). Dotted
    line: 150 km s$^{-1}$, dashed line: 220 km s$^{-1}$, solid line:
    360 km s$^{-1}$. Panel b: SSFR as a function of the stellar mass
    for different redshifts, each symbol corresponds to a specific
    redshift quoted on the plot, the values corresponding to a same
    redshift are connected with a red solid line. Dotted lines
    corresponds to the regression lines proposed for the SAMs of
    \citet{finlator06} (F06) and \citet{daddi07} (D07).}
   \label{predicthiz}
 \end{figure*}

A main prediction issuing from such a secular evolution is the
reduction in the range of values of the SSFR when the redshift
increases (Fig.~\ref{predicthiz}, panel a).  This is a natural result
because for a hypothetical first generation of stars formed during a brief
time $\delta t$, we should have  $SFR ~\times \delta t \sim M_{\rm star}$.
Equivalently the model predicts a flattening in the
variation of the SSFR as a function of the stellar mass as z increases
(Fig.~\ref{predicthiz}, panel b). Such a flattening is indeed observed
in our data up to z=1 and in other studies \citep{iglesias,martin}

At high z, very few observational studies exist. Moreover they are often 
difficult to compare because selection effects are likely to become extremely strong.
For example a K selection at z=4 corresponds to a B rest-frame one.
\citet{feulner} analyzed  the FORS Deep Field and
GOODS-S field to derive the variation in  the SSFR up to z=4 for
different ranges of stellar mass from an I and a K selection. Their
average values for intermediate mass systems ($\rm 10^{9.5}-10^{10.5}
M_{\odot}$) are in reasonable agreement with our models from z=1 to z=3.
In all the redshift range they explore, \citet{feulner} find a strong
decrease in  the SSFR as the mass increases, which might seem at odds
with the predictions of our model (Fig.~\ref{predicthiz}, panels a \&
b). However, \citet{feulner} did not select only star-forming galaxies, 
and the contribution of quiescent objects can steepen the variation in 
the SSFR with the mass since they are essentially very massive
objects. Indeed very recently \citet{daddi07} studied star formation
in massive galaxies forming stars actively up to z$\sim$2 and found a
roughly constant SSFR at z=2 for galaxies detected at 24 $\mu$m and
with a mass comprised between $\sim 5 \, 10^{9}$ and $\rm \sim
10^{11}~M_{\odot}$, in agreement with the predictions of our model.
\citet{daddi07} find an increase in  the SFR at a given mass from z=0
to z=1 and z=2 by a factor $\sim$4 between z=1 and 2 and a factor
$\sim$30 between z=0 and 2, these values are  higher than
those we predict (a factor $\sim$2.5 between z=1 and 2 and a factor
$\sim$10 between z=0 and 2).

\citet{lamareille} present the variation in  the SSFR up to z=2.5 for
the VIMOS VLT Deep Survey in different bins of stellar mass. The
absolute values, as well as the variation they found for the galaxies
with $10<\log(M_{\rm star})<11$, are very consistent with our models. From
their Fig 2 there is also a hint of some flattening of the
SSFR-M$_{\rm star}$ variation when z increases but still of a decrease in 
the SSFR as the mass increases. As in the case of the study of
\citet{feulner}, the VVDS is an I band selected survey that might not
only select star-forming galaxies but also more quiescent objects: a
steep decrease in the SSFR with increasing stellar mass is expected
with  a bimodal galaxy population.

\begin{figure*}
 \centering
  \includegraphics[width=15cm]{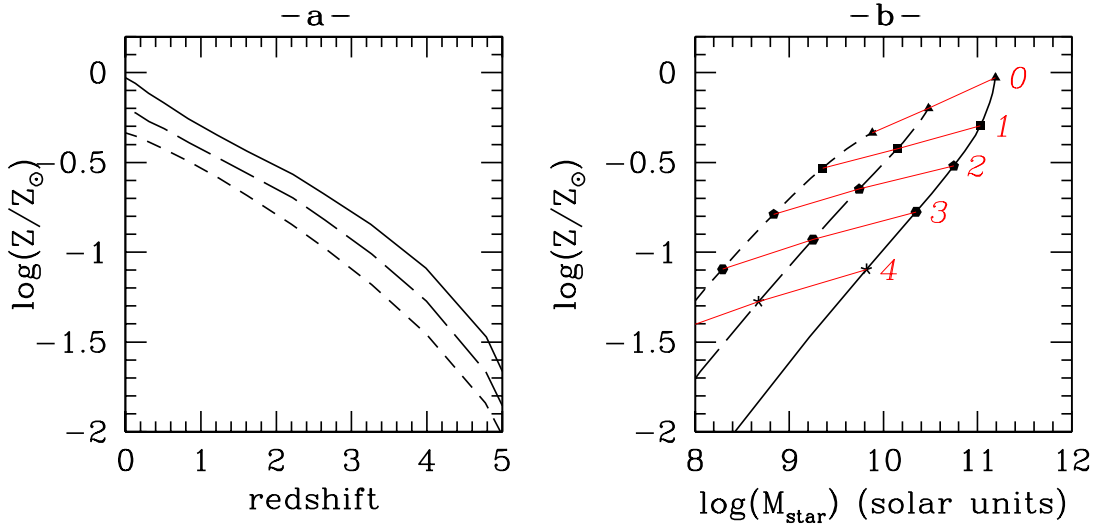}
  \caption{
    Evolution of the metallicity as a function of z (panel a) and of
    the stellar mass (panel d), the symbols and lines are the same as
    in Fig. \ref{predicthiz}}
   \label{predicthiz2}
\end{figure*}

 SAMs have been used in the literature to compare to
   observational sets in the same way we proceeded at z=0.7.
   \citet{daddi07} simulated star-forming galaxies at z=2 from the
   Millennium database. The regression they obtained between the SFR
   and the stellar mass of the simulated galaxies ($\log(SFR) = 0.66
   \log(M_{star})-5.69$) gives SSFRs similar to those obtained with
   our model but with a steeper decrease toward high stellar masses.
   \citet{finlator06} simulated B dropouts at z=4 and also found a
   linear relation between the SFR and the stellar masses: $\log(SFR)
   = 1.14 \log(M_{star})-10.2$. The resulting SSFRs are slightly lower
   than ours at z=4, and they also exhibit a flat distribution of the
   SSFR with the stellar mass. The  relations found by  \citet{daddi07} and \citet{finlator06} are reported in  Fig.~\ref{predicthiz} (panel b).

\subsubsection{Redshift evolution of the metallicity}

Chemical evolution models also allow  predictions of 
the evolution of the metallicity in star-forming galaxies. While 
it is not the case for our sample, recent surveys have allowed to determine
abundances in relatively large numbers of high redshift galaxies. We thus
think it is worth presenting  our models predictions  for 
the evolution of metals.

In our models, we obtain  a progressive increase in the metallicity with time, 
at the same tie as the stellar mass increase with more massive galaxies
always being more metal-rich (as a result of the star formation
histories depending on the mass).  At z=0, a
metallicity-stellar mass relationship is present, corresponding to the
well-known luminosity-metallicity relationship
\citep[e.g. the review by][]{henry99}.  

At higher redshift, we predict that a relation still exists (see Fig. \ref{predicthiz2}), but is
progressively shifted to lower metallicities (by about 0.2 dex at
redshift 1, 0.4 dex at redshift 2). 
We obtain in the models a steepening of the relationship between
redshifts 1 and 0: in the more massive galaxies, the metallicity
increases more rapidly because the gas reservoir is no longer 
replenished by primordial infall, thus the newly formed metals are
less diluted even if the level of star formation is low.

 We note that the obtained change in
metallicity is relatively modest as the theoretical uncertainties
on the models are about a factor 2 (0.3 dex, but this is a systematic effect
that should not affect  our trends but only shift them,  as 
a function of both magnitude and redshift). Another uncertainty comes from the
observational side: it is indeed difficult to estimate abundances,
and various methods produce  different results. For instance,
\citet{liang06}
have shown that the equivalent width method produces systematically
higher abundances by 0.2 dex. It is thus likely that part of the
difference in the observed samples discussed below is  due to the
method used to measure O/H. These variations in metallicity should
then be addressed very carefully.

\citet{rupke07} have recently found that the abundances in LIRGs increased by $\sim$0.2 dex from z$\sim$0.6 to z$\sim$0.1, 
which is roughly consistent with our predictions. Nevertheless, we must remain cautious since the nature of the LIRGs at low and intermediate redshift may well be different \citep[e.g.][]{melbourne}.

There are several studies of the metallicities in high redshift
galaxies, or of the luminosity-metallicity relationships evolution.
However, due to various selection biases and various ways of
estimating the metallicities, they are often difficult to compare with
each other. For this reason, no consensus has been found yet
\citep[][and reference therein]{lamareille06}.  In this 
paper, they investigated the metallicities of 131 intermediate
redshift star-forming galaxies. This allowed them to compare the local
and intermediate redshift (0.2 $<$z$<$ 1), split in 0.2 redshift bins)
mass-metallicity relationship. They did not find any significant
evolution of the slope, but did find that the high redshift
relationship is shifted to lower metallicities (as predicted by our
model) by 0.28 to 0.55 dex (depending on the analysis performed) at z
$\sim$ 1. This is only slightly more  than the value we predicted.

\citet{savaglio05} used the Gemini Deep Deep Survey and the
Canada-France Redshift Survey to investigate the stellar
mass-metallicity relationship between z $\sim$ 0.1 and z $\sim$ 0.7. They
also found that the metallicity is lower at higher redshift for the
same stellar mass by $\sim$ 0.15 dex (according to their figure 13).
They proposed a closed-box model for the chemical evolution of
galaxies able to reproduce this result. This model is based on a simple
exponential decline of the star formation rate after a formation
redshift equal to 3.  The dependence of the exponential folding time
on the total mass of the galaxy is fine-tuned to reproduce the shift
they found in the mass-metallicity relationship (also taking  the metallicities of distant Lyman break galaxies into account).
Such a model is similar in spirit to the one we propose, but ours
is based on a more detailed modelization of the physics of galaxy
evolution (infall, star formation law, finite lifetimes of stars) and was not fine-tuned to 
reproduce the evolution of a simple property between low and high redshifts but
of many properties of nearby galaxies. 

\citet{hammer05} and \citet{liang06} also find that $z \sim$
0.7 emission line galaxies were poorer in metals than present-day
spirals, by 0.3 dex, a result also consistent with the average
  trend found by \citet{maier05}. 
 On the other hand, \citet{kobulnicky04} report a smaller 
variation in 0.14 dex from $z=0$ to $z=1$.
The factor two difference with other studies might be linked to the fact that they used equivalent widths, and standard underlying stellar absorption
rather than using high quality calibrated spectra and measuring the Balmer absorption.
Their data suggest a
steepening of the slope of the metallicity-uminosity relationship.
One should also note that a direct comparison in this case is
difficult since the models show the metallicity-stellar mass
relationship. While luminosity scales with the mass in nearby normal
galaxies, the luminosity of high redshift galaxies might be strongly
affected by their current star formation rate.  Going further,
\citet{erb06} suggests that star-forming galaxies at redshift $\sim$ 2
have 0.3 dex fainter metallicities. This is a somewhat smaller
decrease than expected from our models;  but taking into account the
issues of selection and metallicity calibration, it is (at least) not
inconsistent with them.

Because of the uncertainties in measuring metallicities and in
converting stellar masses to B band luminosities, a more detailed
comparison is quite pointless. We hope that Fig.~\ref{predicthiz2}
gives a sketch of the evolutionary trends expected in the case of
secular evolution, from which eventual departures should be due to
more complex physics  \citep{rupke07}.  Determinations at high redshift of star
formation rates, stellar masses, and metallicities may be compared to
these trends to know how much galaxies are, on average, far from a
secular evolution or not at each redshift.
For instance, the values of stellar masses and metallicities
in $z=5$ Lyman break galaxies of \citet{ando07} are incompatible
with  simple secular evolution objects. It is thus likely that these
objects are not precursors of normal nearby spirals.

\section{Conclusions}
   
We have analyzed the star formation rate and the stellar mass of
galaxies selected to be active in star formation at z=0 and z=0.7. The
selection was  performed in UV and in IR (rest frame). As long as
relatively massive systems are studied ($M_{\rm star}>10^{10} {\rm M_{\odot}}$),
the IR selection is found to be more efficient than the UV one to select all
the star-forming systems. Nevertheless the galaxies selected in UV and
in IR exhibit similar variations in the SSFR. 

We compared mean relationships between the observed specific star
formation rate and the stellar mass  at z=0 and z=0.7 
with physically motivated models aimed at
reproducing the mean properties of local spiral galaxies and of the Milky
Way. These models are based on a progressive infall of
gas into the galactic disk starting at high z. The
agreement is found to be quite good given the uncertainties in the
models and the data. Both data and models exhibit a fairly flat
distribution of SSFR for galaxies with masses  between
$\rm 10^{10} M_{\odot}$ and $\rm 10^{11} M_{\odot}$,   this flattening being more
pronounced at z=0.7 than at z=0. These results are  consistent with those
obtained at intermediate  redshift from surveys selecting
star-forming galaxies.

We have proposed an analytical formula for the mean evolution of the SFR
with time  and the  parameters vary with the current stellar mass of the
galaxy. This formulation will allow anyone to perform
very simple calculations predicting e.g. star formation rates, stellar
masses, or metallicities.  We present predictions for the values of
the specific star formation rate, stellar mass, and metallicity at high
redshift.
These predictions can be taken as templates for a secular evolution of
galaxies, including gas accretion. They are found consistent with the mean trends deduced from the simulations of semi-analytical models up to z=4.
We  tentatively compared these predictions to  some  existing data
concerning star-forming galaxies up to z$\sim$2 without finding major
departures. 

The comparison of model predictions with observations is
difficult because of uncertainties (in observations, models, star
formation rate, or metallicity calibrations) and selection biases.  
Thus we wish to stress that the trends presented in Fig
\ref{predicthiz} are only indications of the expected mean evolution in a
simple scenario. Caution should be taken when comparing them to
data. However, we believe they will still be useful for comparisons with the
results of future or  on going large surveys, especially if star
formation rates, stellar masses, and metallicities are computed in
consistent ways at different redshifts.

\begin{acknowledgements}

TTT has been supported  by Program for Improvement of Research Environment for 
Young Researchers from Special Coordination Funds for Promoting Science 
and Technology commissioned by the Ministry of Education, Culture, Sports,
Science and Technology (MEXT) of Japan.  ELF acknowledges the support from the
Spitzer Space Telescope Fellowship Program through a contract issued by JPL/Caltech and NASA.
ME acknowledges the suport from the Tokyo Keizai University Research
Grant (A07-05). MN was supported by the Grant-in-Aid for the Scientific Research Fund
(18749007) of the Ministry of Education, Culture, Sports, Science and
Technology of Japan and by a Nagasaki University president's Fund
grant.

\end{acknowledgements}

\end{document}